\documentstyle{llncs}
\newtheorem{defn}{D\'efinition}[section]

\newtheorem{thm}[defn]{Theorem}
\newtheorem{lem}[defn]{Lemma}

\begin{document}
\title{PRECISE ESTIMATION ON THE ORDER OF LOCAL TESTABILITY OF 
DETERMINISTIC FINITE AUTOMATON}
\author{A.N. Trahtman} 
\date{1998}
\institute{Bar-Ilan University, Department of Mathematics and Computer Science,
 52900 Ramat Gan,Israel, {\it e-mail}: trakht@macs.biu.ac.il}

\maketitle

\centerline{Lectures Notes of Computer Sciences, 1436(1998), 198-212}
\centerline{AUTOMATA IMPLEMENATION}

\medskip

\begin{abstract} A locally testable language $L$ is a language with
        the property that for some nonnegative integer  $k$, called the order
        or the level of local testability, whether or not a word $u$ in the
	language $L$ depends on (1) the prefix and
        suffix of the word $u$ of length $k-1$ and (2) the set of
        intermediate
        substrings of length $k$ of the word $u$.
        For given $k$ the language is called $k$-testable.

 We give necessary and sufficient conditions for the language of an
 automaton to be $k$-testable in the terms of the length of paths of a related
  graph.
Some estimations of the upper and of the lower bound of order of testability
 follow from these results.

   We improve the upper bound on the order of testability of locally
 testable deterministic finite automaton with $n$ states  to
 ${n^2-n \over 2}+1$. This bound is the best possible.

   We give an answer on the following conjecture of Kim, McNaughton and
McCLoskey for deterministic finite locally testable  automaton with $n$ states:
``Is the order of local testability no greater than $\Omega(n^{1.5})$ when the
alphabet size is two?''

Our answer is negative. In the case of size two the situation is the same
as in general case: the order of local testability is $\Omega(n^2)$.
\end{abstract}


 \section{Introduction}.

   The concept of local testability was first introduced by McNaughton
 and Papert $\cite {MP}$ and since then has been extensively investigated from
 different points of view (see  $\cite {BS}$,  $\cite {C}$,
 $\cite {K89}$ - $\cite {K94}$,  $\cite {Ma}$, $\cite {Pa}$,  $\cite {Pi}$,
 $\cite {Tr}$, $\cite {Z}$, $\cite {ZC}$). This concept is connected with
 languages,
 finite automata and semigroups.
 In $\cite {Mi}$, local testability
 is discussed in terms of "diameter-limited perceptrons". Locally testable
languages are a generalization of the definite and reverse-definite
languages, which can be found, for example, in $\cite {G}$ and $\cite {Sh}$.

  In  $\cite {K91}$
   necessary and sufficient conditions for an automaton to be locally
testable were found. In  $\cite {K94}$ the NP-hardness of finding of the
order of local testability was proved. The necessary and sufficient
conditions of $k$-testability in the terms of 5-tuple graph were found in
$\cite {K94}$.  An estimation for the order of local testability for an
arbitrary deterministic finite automaton was found first in  $\cite {K91}$
  and then improved in  $\cite {K94}$. The upper bound from  $\cite {K94}$ is
 $2n^2+1$, where $n$ is the number of states of the automaton.

   For the state transition graph $\Gamma$ of an automaton we consider some
subgraphs of the direct product $\Gamma \times \Gamma$.
   We introduce in this paper  sufficient and necessary conditions
 for the automaton and transition
 semigroup  of the automaton to be $k$-testable in terms of the length of some
 paths without loops on these graphs. This gives us some upper and some lower
 bounds on the order of local testability.

 In the case that the state transition graph is strongly connected the
 sufficient conditions are necessary as well and algorithm of finding of the
level
of local testability is polynomial and not NP-hard as in the general case
 $\cite {K94}$.

 As corollary we receive
the precise upper bound on the order of local testability
for deterministic finite locally testable reduced automaton with $n$
 states.
 It is equal to $(n^2-n)/2+1.$ This result improves the estimations from
 $\cite {K91}$,  $\cite {K94}$ and finishes investigations in this direction.

 In $\cite {K89}$ and  $\cite {K94}$ one can find conjecture
that in the case of the alphabet two
the upper bound on the order of local testability
for the deterministic finite locally testable reduced automaton with $n$
 states is not greater than $\Omega(n^{1.5})$.

We consider in this paper an  example of sequence of deterministic finite
 automata with
 $n$ states whose alphabet size is two. It will be proved that the considered
automata are
locally testable and their order of local testability is $\Omega(n^2)$. So the
problem from $\cite {K89}$, $\cite {K94}$  is solved negatively.

 Our example is one between examples of locally testable automata
whose order of testability is greater  than the number of its states.
First such astonishing example of an automaton with 28 states had appeared in
 $\cite {K89}$ -  $\cite {K94}$.
(Note that the order of testability of the
considered automaton found in these papers is
not correct.  It is more greater than 126 $\cite {K89}$  or 127  $\cite
 {K94}$.  The conjuncture of the authors that the automaton has the maximal
 order of testability for automata with 28 states and alphabet size two
is not correct too. There
exist a deterministic finite 142-testable automaton with 28 states and
alphabet size two).

 The description of the identities of $k$-testable semigroup
 from $\cite {Tr}$ is used here. The concept of the graph is
 inspired by the works $\cite {K89}$, $\cite {K91}$ 
 of Kim, McNaughton and McCLoskey.
  The purely algebraic approach proved to be
 fruitful (see $\cite {Pa}$, $\cite {Tr}$, $\cite {Z}$) and
in this paper we use this technique too.
  The results of the work are announced in $\cite {TW}$.

 \section{Notation and definition}

 Let $\Sigma$ be an alphabet and let $\Sigma^+$ denote  the free semigroup
on $\Sigma$. If $w \in \Sigma^+$, let $|w|$ denote the length of $w$.
Let $k$ be a positive integer. Let $i_k(w)  [t_k(w)]$ denote the prefix
[suffix] of $w$ of length $k$ or $w$ if $|w| < k$. Let $F_k(w)$ denote  the
 set of factors of $w$ of length $k$.
 A language $L$ [a semigroup $S$] is called {\it k-testable} if there is an
 alphabet $\Sigma$
[and a surjective morphism $\phi : \Sigma^+ \to S$] such that  for all
$u$, $v \in \Sigma^+$, if $i_{k-1}(u)=i_{k-1}(v), t_{k-1}(u)=t_{k-1}(v)$ and
 $F_k(u)=F_k(v)$, then either both $u$ and $v$ are in $L$ or neither
 is in $L$ [$u\phi = v\phi$].

   This definition follows  $\cite {BS}$, $\cite {K89}$.
 In $\cite {MP}$ the definition differs by considering prefixes and suffixes of
 length $k$.

   An automaton is $k$-testable if the automaton accepts a $k$-testable
 language [the syntactic semigroup of the automaton is $k$-testable].

   A language $L$   [a semigroup $S$, an automaton $\bf A$] is {\it locally
 testable} if it is $k$-testable for some $k$.

 For local testability the two definitions mentioned above are equivalent
 $\cite {K89}$ .

   It is known that the set of $k$-testable semigroups forms a variety of
 semigroups ($\cite {L}$, $\cite{Z}$).
   Let $T_k$ be the variety of $k$-testable semigroups.

 $|S|$ - the number of elements of the set $S$.

 $|d|$ - the length of the word $d$ in some alphabet.

   $S^m$  - the ideal of the semigroup $S$ containing products of elements of
   $S$ of length $m$ and
   greater.

   We say that the element $a$ from a semigroup $S$ divides the element $b$
 from $S$ if $b=dac$ for some $c, d \in S \cup \emptyset$.

  According to  the result from $\cite {Tr}$ $T_n$ has the following
 basis of identities:

 \begin{equation}
 \alpha_r: (x_1 ... x_r)^{m+1}x_1...x_p=(x_1...x_r)^{m+2}x_1...x_p \label{(1)}
 \end{equation}

 where $r \in \{1,...n\}$, $ p=n-1$(mod $r$), $m=(n-p-1)/r$, $n=mr+p+1$,
 \begin{equation}
 \beta: x_1...x_{n-1}yx_1...x_{n-1}zx_1...x_{n-1}=%
 x_1...x_{n-1}zx_1...x_{n-1}yx_1...x_{n-1}   \label{(2)}
\end{equation}

   For instance, $\alpha_1: x^n=x^{n+1}$. A locally testable semigroup
$S$ has only trivial subgroups $\cite {BS}$ and so a locally testable
semigroup $S$ with $n$ elements  satisfies  identity $\alpha_1$.

A maximal strongly connected component of the graph will be called
 $SCC$ $\cite {K89}$

 Let $\Gamma$ be the state transition graph of a finite automaton with
 edges labeled by elements of $\Sigma$.

  The state transition graph $\Gamma$ of a finite automaton is
called {\it complete} if for every node $\bf p$ $\in \Gamma$ and
every $\sigma \in \Sigma$ we have ${\bf p}\sigma \in \Gamma$.
  Any state transition graph $\Gamma$ of a finite automaton may be
transformed in complete graph by adding sink state.

 The element $e \in \Sigma^+ $ ($\in S$) will be called {\it right unit}
 of the node $\bf p$ $\in \Gamma$ if $\bf p$$e=\bf p$.

 We shall write $\bf p \succeq \bf q$ if the node $\bf q$  is reachable from
the node $\bf p$ and $\bf p \succ \bf q$  if $\bf p \succeq \bf q$  and the
nodes $\bf p, q$ are distinct.

In the case $\bf p \succeq q$ and $\bf q \succeq p$ we write $\bf p \sim q$
($\bf p$ and $\bf q$ belong to one $SCC$).

 We construct now a edge-labeled directed graph $\Gamma\Gamma$ on the nodes
 ($\bf p, \bf q$) where ${\bf p, q} \in \Gamma$ and $\bf p \succ \bf q$.
We say $(\bf {p,q}) \to (\bf {r,t})$ iff for some $\sigma \in \Sigma$ we have
 ${\bf p} \sigma=\bf r$ and ${\bf q} \sigma=\bf t$.  The corresponding edge in
$\Gamma\Gamma$ will be labeled by $\sigma$.  The graph $\Gamma\Gamma$ will be
called the {\it 2-tuple} graph of the automaton.

The path $\Phi$ on the 2-tuple graph $\Gamma\Gamma$ will be called {\it
SCC-restricted} if all components of its nodes belong to one $SCC$ of $\Gamma$.

Consider a path $\Phi$: (${\bf p}_1,{\bf q}_1$),...(${\bf p}_k, {\bf q}_k$)
on the 2-tuple graph for which there exist $\sigma \in \Sigma$ such that
${\bf p}_k\sigma \not\succ {\bf q}_k\sigma$ and
 ${\bf q}_k\sigma \succ {\bf q}_1$ on the graph $\Gamma$. Note that all
 $\bf q_i$
belong to one $SCC$ of $\Gamma$.  The path $\Phi$ will be called a {\it
SCC-semirestricted path}.

Consider a path $\Phi$ on the graph $\Gamma\Gamma$ with the nodes
(${\bf a}_1,{\bf b}_1$), (${\bf a}_2,{\bf b}_2$),... (${\bf a}_s,{\bf b}_s$)
 such that
there exist a natural number $r$ such that ${\bf a}_{i+r}={\bf b}_i$ for all
  possible natural $i$ and for each $j$ there exist such $\sigma \in \Sigma$
  that for all $i \geq 0$ we have
  $({\bf a}_{j+ri},{\bf b}_{j+ri})\sigma=({\bf a}_{j+ri+1},{\bf b}_{j+ri+1})$.
  The  path $\Phi$ will be called {\it $r$-periodic path}.

A  path without loops is called $\it simple$.
 A path without common nodes with any $SCC$ will be called {\it strongly
simple}.

The {\it length of a path} is the number of edges on the path.
  \medskip
 \section{The graph of the automaton}

  We present two key lemmas of Kim, McNaughton and McCLoskey
in the following convenient form:
 \begin{lem} ( $\cite {K89}$)
  Let the nodes
  $\bf p, q$ belong to one $SCC$ of the state transition graph of a
      locally testable deterministic finite automaton.

 Then the node
 ( $\bf p,q$) does not belong to some $SCC$ of the  2-tuple graph
  $\Gamma\Gamma$ of the automaton.
   \end{lem}.

  \begin{lem} $\label {3.1a}$ ( $\cite {K91}$, Lemma 4) Let the node
 ( $\bf p,q$) belong to some $SCC$ of the  2-tuple graph $\Gamma\Gamma$  of
  a locally  testable deterministic finite automaton and let $s$ be an
  arbitrary element of the transition semigroup of the automaton.

  Then  $\bf p$$s \succeq \bf q$ is valid iff $\bf q$$s \succeq \bf q$
 on the state transition graph of the automaton.
   \end{lem}.

   Both these lemmas give us
necessary and sufficient conditions for a deterministic finite automaton to
be locally testable $\cite {K89}$, $\cite {K91}$.

\begin{lem} $\label {3.2}$
 Let $S$ be transition semigroup of a locally testable reduced
deterministic finite automaton and let $\Gamma\Gamma$ be its 2-tuple graph.
Suppose for some elements $a_1,...a_r \in S$ for some nonnegative $m$
and $ p<r $ we have
$(a_1...a_r)^{m+1}a_1...a_p \neq (a_1...a_r)^{m+2}a_1...a_p $.

 Then on the graph $\Gamma\Gamma$ there exist a simple path
of the length $mr+p $.

 If on the graph $\Gamma\Gamma$ there is no simple path of
 length $k-1$ then the identities $\alpha_r$ (1) of $k$-testability are valid
 on $S$.
\end{lem}

 Proof. It follows from the given inequality that for some node $\bf q$ from
the state transition graph $\Gamma$ we have
$\bf q$$(a_1...a_r)^{m+1}a_1...a_p \neq \bf q$$(a_1...a_r)^{m+2}a_1...a_p $.
At least one of two parts of the inequality is a node of $\Gamma$.
It implies that $\bf q$$(a_1...a_r)^{m+1}a_1...a_p $ is a node of $\Gamma$.
Denote the left subword of the word $(a_1...a_r)^n$ of length $i$ by $b_i$.
On the graph $\Gamma\Gamma$ there exist a path $\phi$ from the node
 $(\bf q, \bf q$$b_r)$
to the node $(\bf q$$b_i, \bf q$$b_{i+r})$ and its minimum length is $mr+p $.
Our aim is now  to find on this path a simple subpath of the
 necessary length.

  So suppose that $\phi$ is not simple and there exist a loop on the path
 $\phi$.
 Let the nodes on the places $k$ and $k+j$ coincide for the first such loop
 from the left.  So $\bf q$$b_k=\bf q$$b_{k+j}$ and
$\bf q$$b_{k+r}=\bf q$$b_{k+r+j}$.  Then the two nodes $\bf q$$b_k$ and
$\bf q$$b_{k+r}$ from $\Gamma$ have the same right unit.  In view of lemma 3.1
the nodes $\bf q$$b_k$ and $\bf q$$b_{k+r}$ belong to different $SCC$.  From
lemma $\ref {3.1a}$ it follows that all nodes $\bf q$$b_l$ for $l \geq k+r$
on the considered path $\phi$ belong to the same $SCC$ of $\Gamma$.  If the
node $\bf q$$b_{mr+r+p}$ exists, then the node ${\bf q}b_{mr+2r+p}$ exist as
well.  After the node $({\bf q}b_{k+r}, {\bf q}b_{k+2r})$ on $\phi$ there are
no loops (Lemma 3.1).  Hence, $j<r$. There are no loops on the path before
node $({\bf q}b_k, {\bf q}b_{k+r})$ by the choice of $k$ . We can exclude all
possible loops between these two nodes and obtain a subpath without loops.

  From the existence of node $\bf q$$b_{mr+2r+p} \in \Gamma$
 it follows that the
length of the path $\phi$ is $(m+1)r+p $ and the length of this simple
subpath will be at least $mr+p+1$.

  In the case there are no loops on $\phi$, the length of $\phi$
 will be at least $mr+p $. This follows from existence of the node
 $\bf q$$b_{mr+r+p}$.

  The first part of the statement of lemma is proved.

  Suppose now that on $\Gamma\Gamma$ there are no simple paths of
the length $k-1$.
Then for $k-1<mr+p $ and for any $\bf q$ $ \in \Gamma$ we have
${\bf q}(a_1...a_r)^{m+1}a_1...a_p={\bf q}(a_1...a_r)^{m+2}a_1...a_p $.
  The second statement of the lemma follows now from the first and from the
description of the identities (1) of $k$-testability.
\medskip

\begin{lem} $\label {3.3}$
 If on the 2-tuple graph $\Gamma\Gamma$ of a deterministic finite
automaton there exist an $r$-periodic path of length $k+r-1$ then the
 automaton is not $k$-testable.

$k+1$ is a lower bound on the order of local testability of the automaton

\end{lem}

  Proof. Suppose that the automaton is locally testable.
 Let ($\bf s, q$) be the first node on the considered $r$-periodic path and
elements $a_1, ... a_r$ from $\Sigma$ denote the first  $r$ edges of the path.
So ${\bf s}a_1...a_r=\bf q$ and the last node on the path is
(${\bf s}(a_1...a_r)^{m+1}a_1...a_p, {\bf s}(a_1...a_r)^{m+2}a_1...a_p $) where
$mr+p=k-1$, $p<r$, $m \geq 0$. Number of the edges on the path is
 $(m+1)r+p=k+r-1$.
The components of the nodes are distinct and so the existence of the last node
 proves that $(a_1...a_r)^{m+1}a_1...a_p \neq (a_1...a_r)^{m+2}a_1...a_p $.
Then the identity $\alpha_r$
from (1) for $k$-testability is not valid on the transition semigroup of the
automaton.

  The lemma is proved.

\begin{lem} $\label {3.4}$
 Suppose that on the 2-tuple graph  of a deterministic finite
 locally testable automaton there exist an $SCC$-restricted path of
length $k-1$.

  Then the identity $\beta$ (2) of $k$-testability is not valid on the
 transition semigroup $S$ of the automaton and both $S$ and the automaton are
not $k$-testable.

$k+1$ is a lower bound on the order of local testability of the automaton
 \end{lem}

  Proof. In order to prove the non-validity of the identity $\beta$ we must
  find elements $a_1,..., a_{k-1}$, $b$, $c$ $\in S$  such that

\begin{equation}
 a_1...a_{k-1}ba_1...a_{k-1}ca_1...a_{k-1} \neq a_1...a_{k-1}ca_1...a_{k-1}ba_1...a_{k-1}
\end{equation}
Let $(\bf s, q)$ be the first node on the considered path on
 $\Gamma\Gamma$ and elements $a_1,..., a_{k-1} \in S$ denote edges of the
path.  Let us denote $a=a_1...a_{k-1}$.  So the node (${\bf s}a,{\bf q}a$)
 is the last node on the path.  Hence, ${\bf s}a \neq {\bf q}a$ and the nodes
${\bf s, q,s}a,{\bf q}a$
belong to one $SCC X$ of $\Gamma$. So there are elements  $b, c \in S$ such
that ${\bf s}ab=\bf s$, ${\bf s}ac=\bf q$.  Without loss of generality let us
assume that element $b$ is divided by an idempotent $e \in S$. This follows
from the equality $s(ab)^n=s$ and local testability of $S$.  Thus,
$b=b_1eb_2$ for some $b_1, b_2 \in S$.

 If ${\bf s}ab_1e \not\sim {\bf q}ab_1e$
from the fact that ${\bf s}abaca={\bf s}ab_1eb_2aca={\bf q}a \in X$ it
follows that  ${\bf s}ab_1e \in X$ and ${\bf q}ab_1e \not\in X$ because
distinct $SCC$ are not connected with a loop.

Note that for any node ($\bf p, r$) such that the nodes
 $\bf p,r$ lie  outside
the $SCC$ $X$ and are reachable from $X$ and for any node
$(\bf s,\bf t)$ such that
${\bf s, t} \in X$ we have ($(\bf p, r) \not\succ (\bf s, t)$).

 The node
${\bf q}ab_1e$ lies outside $X$, then ${\bf s}acaba={\bf q}aba$ does not
 belong to $X$ too.  From ${\bf q}a \in X$ we have ${\bf q}a={\bf s}abaca
\neq {\bf s}acaba={\bf q}aba$, whence $abaca \neq acaba$.

  So we may suppose that ${\bf s}ab_1e \sim {\bf q}ab_1e$   and
 ${\bf q}ab_1e \in X$. The nodes ${\bf q}ab_1e$ and
 ${\bf s}ab_1e$ have the common right unit $e$ and belong to the same $SCC$.
 From lemma 3.1 it follows that ${\bf s}ab_1e={\bf q}ab_1e$. Then
  ${\bf s}ab={\bf q}ab$ and
 ${\bf q}ab=\bf s$.  This implies ${\bf s}acaba={\bf q}aba={\bf s}a$. Now from
 $\bf s$$abaca=\bf s$$aca=$$\bf q$$a$ and ${\bf q}a \neq {\bf s}a$ it
 follows that ${\bf s}acaba \neq {\bf s}abaca$ and $abaca \neq acaba$.

  The lemma is proved. \medskip

\begin{lem} $\label {3.5}$
 Suppose that on the 2-tuple graph $\Gamma\Gamma$ of a
deterministic finite locally testable automaton with state transition graph
$\Gamma$ there exist $SCC$-semirestricted path $\phi$.

Then the second components of all nodes of the path $\phi$ belong to one
$SCC$ of $\Gamma$ and no node of the path $\phi$ does not belong to some
 $SCC$ of the 2-tuple graph $\Gamma\Gamma$.
\end{lem}

 Proof. For the first node (${\bf p}_1, {\bf q}_1$) and the last node
(${\bf p}_i, {\bf q}_i$) of the path $\phi$ we have
 $({\bf p}_1, {\bf q}_1) \succ ({\bf p}_i, {\bf q}_i)$ and ${\bf q}_i \succ {\bf q}_1$.
Hence, ${\bf q}_1 \sim {\bf q}_i \sim {\bf q}_j $ for any $j<i$.

 Suppose that the considered path $\phi$ has a common node ($\bf p, \bf q$)
with some $SCC$ of $\Gamma\Gamma$. Then for some element $e$ from transition
semigroup $S$ we have ${\bf p}e=\bf p$, ${\bf q}e=\bf q$, $\bf p \succ \bf q$.
Then the necessary condition of local testability (lemma $\ref {3.1a}$)
 implies that
for any $x \in S$ such that ${\bf q}x \succ {\bf q}$ we have
${\bf p}x \succ {\bf q}x$. Therefore  the node ($\bf p, \bf q$)
 could not belong to an $SCC$-semirestricted path.

 \begin{lem} $\label {3.6}$
Suppose that on the 2-tuple graph $\Gamma\Gamma$ of a
deterministic finite automaton with state transition graph $\Gamma$ there exist
SCC-semirestricted path of length $k-1$.

Then the identity $\beta$ of $k$-testability is not valid on the transition
semigroup $S$ of the automaton and both $S$ and the automaton are not
$k$-testable.

$k+1$ is a lower bound on the order of local testability of the automaton.
\end{lem}

  Proof. In order to prove the non-validity of identity $\beta$ we must find
elements $a_1,...,a_{k-1}$, $b$, $c \in S$ such that for $a=a_1...a_{k-1}$
we have $abaca \neq acaba$ (See (3)).

  Let $a_1,...,a_{k-1}$ denote the edges of the considered path
(${\bf p}_1,{\bf q}_1$),...,(${\bf p}_k,{\bf q}_k$) and $a=a_1...a_{k-1}$.
Suppose that ${\bf p}_k\sigma \not\succ {\bf q}_k\sigma$ on $\Gamma$ for
some $\sigma \in \Sigma$ such that ${\bf q}_k\sigma \succ {\bf q}_1$ .
 From the preceding lemma and the definition of $SCC$-semirestricted
path it follows that the nodes ${\bf q}_1$, ${\bf q}_k$ and ${\bf q}_k\sigma$
belong to one $SCC$ of $\Gamma$ and ${\bf p}_k \succ {\bf q}_k$, whence there
exist an element $b \in S$ such that ${\bf p}_kb={\bf q}_1$. By the
above-mentioned definition there exist an element $c=\sigma d \in S$ such
that ${\bf q}_kc={\bf q}_1$.  Then
 ${\bf p}_1abaca={\bf p}_kbaca={\bf q}_1aca={\bf q}_kca={\bf q}_1a={\bf q}_k$.
  Consider the
node ${\bf p}_1acaba={\bf p}_k\sigma daba$. The node ${\bf q}_k\sigma$ is not
reachable from ${\bf p}_k\sigma$ and so ${\bf p}_k\sigma \not\succ {\bf q}_k$,
whence ${\bf p}_k\sigma daba \neq {\bf q}_k$.  
 So ${\bf p}_1abaca \neq {\bf p}_1acaba$ and $abaca \neq acaba$.

  The lemma is proved.
\medskip

\begin{lem} $\label {3.7}$
Let $S$ be the transition semigroup of a locally testable
reduced deterministic finite automaton and suppose that
on the 2-tuple graph $\Gamma\Gamma$ of the automaton
there are no strongly simple paths of length $k-1$. Suppose that $x \in
 S^{k-1}$, $y, z \in S$ and $S$ satisfies the identity $xyx=xyxyx$

 Then $S$ satisfies identity $xyxzx=xzxyx$. (identity $\beta$ for
 $k$-testability)

\end{lem}

    Proof. From the identity $xyx=xyxyx$ we deduce the following identities
 \begin{equation}
 xzx=xzxzx,
 xzxyx=xzxyxyx,
 xzxyx=xzxyxzxyx  \label{(3)}
\end{equation}
 for $x \in S^{k-1}$, $y,z \in S$.
  So the words $xyxzx, xzxyx, xzxyxy, xyxzxz$ divide each other
 in $S$.

  Let us suppose that the identity $xyxzx=xzxyx$ is not valid on $S$.
  Then for some node ${\bf p} \in \Gamma$ and for some
 $x \in S^{k-1}, y, z \in S$ we
 have ${\bf p}xzxyx \neq {\bf p}xyxzx$. Without loss of generality
let us assume that there exists a node ${\bf p}xyxzx$.

Suppose first that ${\bf p}x \neq {\bf p}xyxzx$.
 Consider the path  from the node (${\bf p}, {\bf p}xyxz$) to the node
( ${\bf p}x, {\bf p}xyxzx$) in $\Gamma\Gamma$. In view of $|x| \geq k-1$
some node on the path belongs to an $SCC$. The element $x$ may be presented
in the form $x_1x_2$ such that the nodes ${\bf p}x_1$ and
 ${\bf p}xyxzx_1$ have
a right unit in $\Gamma$. Now from the necessary condition of local
testability (Lemma $\ref {3.1a}$) it follows that
 ${\bf p}xs \succeq {\bf p}xyxzxs$ in $\Gamma$ for any $s \in S$
 such that $xs$ is a left subword of
the word of (4).  Let $s=zxyx$.  Then
 ${\bf p}xzxyx \succeq {\bf p}xyxzxzxyx={\bf p}xyxzxyx$.

   The equality  ${\bf p}xyxzx={\bf p}xyxzxyxzx$ follows from (4) and it
 implies
 that the nodes ${\bf p}xyxzx$ and ${\bf p}xyxzxyx$ belong to the same $SCC$ of
$\Gamma$.  Then in $\Gamma$  ${\bf p}xzxyx \succeq {\bf p}xyxzx$ and the first
node of the formula exists.

In the case that ${\bf p}x={\bf p}xyxzx$ we have ${\bf p}xzx={\bf p}xyxzx$ and
${\bf p}xzx={\bf p}xyxzx={\bf p}x$. So ${\bf p}x={\bf p}xzxyxzx$.
 Hence ${\bf p}xzxyx \succeq {\bf p}xyxzx$  and the
 node ${\bf p}xzxyx$ exist as well.

Now from the existence of the node ${\bf p}xzxyx$ it follows in analogous way
 that  ${\bf p}xyxzx \succ {\bf p}xzxyx$.
   Thus,
 both nodes ${\bf p}xyxzx$ and ${\bf p}xzxyx$ belong to the same $SCC$.

  The nodes ${\bf p}xyxzxz$ and ${\bf p}xzxyxy$ belong to the same $SCC$
 as well.
Multiplying by $x$ the nodes of one $SCC$ must unite them because the result
belong to the same $SCC$, $|x| \geq k-1$ and on the path corresponding to
$x$ there are no loops.

 So ${\bf p}xyxzxzx={\bf p}xzxyxyx$ for every ${\bf p} \in \Gamma$.
 Thus, $S$ satisfies the identity $xzxyxyx=xyxzxzx$.
In view of the identity $xyx=xyxyx$ we get that $xyxzx=xzxyx$.

 The lemma is proved.

\proclaim Corollary. Let $S$ be the transition semigroup of a locally testable
reduced deterministic finite automaton and suppose that
on the 2-tuple graph $\Gamma\Gamma$ of the automaton
there are no simple paths of length $k-1$. Suppose that $x \in S^{k-1}$,
 $y, z \in S$ and $S$ satisfies the identity $xyx=xyxyx$.

  Then $S$ satisfies identity $xyxzx=xzxyx$. (identity $\beta$ for
 $k$-testability).
  \medskip

\begin{lem} $\label {3.8}$
Let $S$ be the transition semigroup of a locally testable reduced
deterministic finite automaton and suppose that on the 2-tuple graph
 $\Gamma\Gamma$
of the automaton there are no $SCC$-restricted paths of length $k-1$.
Suppose that $x \in S^{k-1}, y, z \in S$ and $S$ satisfies the identity
$xyx=xyxyx$.

Then $S$ satisfies the identity $xyxzx=xyxzxyx$.
\end{lem}

Proof. From the identity $xyx=xyxyx$ follow identities (4). This implies
 that
the words $xyxzx$, $xyxzxyx$, $xyxzxyxy$, $xyxzxz$ are divided one by another.
So the nodes ${\bf p}xyxzx$, ${\bf p}xyxzxz$, ${\bf p}xyxzxyx$,
 ${\bf p}xyxzxyxy$
belong to a common $SCC$ of $\Gamma$. Suppose that
 ${\bf p}xyxzxzx \neq
{\bf p}xyxzxyxyx$. Then on the 2-tuple graph $\Gamma\Gamma$ there exists a
path from the node (${\bf p}xyxzxz, {\bf p}xyxzxyxy$) to the node
 (${\bf p}xyxzxzx, {\bf p}xyxzxyxyx$).
 We obtain a $SCC$-restricted path of the length $|x|=k-1$
 This contradicts our assumption. So ${\bf p}xyxzxzx={\bf p}xyxzxyxyx$.
 In view
of (4) we have ${\bf p}xyxzx={\bf p}xyxzxyx$.

The node $\bf p$ is an arbitrary node and so $xyxzx=xyxzxyx$.

  The lemma is proved.
\medskip

 \begin{thm} $\label {3.9}$
  Let $S$ be the transition semigroup of a reduced deterministic finite
locally testable automaton {\bf A} and $\Gamma\Gamma$ its 2-tuple graph.
 Assume the graph $\Gamma\Gamma$ does not contain simple
paths of length $k-1$.

 Then both the automaton {\bf A} and the semigroup $S$ are $k$-testable.

 $k$ is an upper bound on the order of local testability of the automaton
\end{thm}

 Proof. The validity of the identities $\alpha_r$
for $k$-testability follows from lemma $\ref {3.2}$.
The validity of the identity $\beta$ in view of validity of $\alpha_k$
follows from corollary of lemma $\ref {3.7}$.
\medskip.

  From theorem $\ref {3.9}$ and lemmas $\ref {3.4}$, $\ref {3.6}$ we
 immediately obtain the following result.

\begin{thm} $\label {3.10}$
Let $\Gamma\Gamma$ be the 2-tuple graph of a locally testable
deterministic reduced finite automaton {\bf A}. Let the maximum
length of $SCC$-restricted and $SCC$-semirestricted paths on
 $\Gamma\Gamma$ be equal to $k-2$.

Then the identity $\beta$ of ($k-1$)-testability is not valid on the
transition semigroup $S$ of the automaton {\bf A} and both $S$ and the
automaton are not ($k-1$)-testable, $k$ is a lower bound on the order of
local testability.

 If the length of all simple  paths on $\Gamma\Gamma$ is
not greater than $k-2$ then {\bf A} is precisely $k$-testable.
 \end{thm}
\medskip
\begin{thm} $\label {3.11}$
Assume that the state transition graph $\Gamma$ of a locally
 testable reduced
deterministic finite automaton {\bf A} is strongly connected. Let the maximum
 of the lengths of strongly simple [simple] paths on the 2-tuple graph of
{\bf A} be $k-2$.

Then the automaton is precisely $k$-testable.
\end{thm}

 The proof follows from the preceding theorem and from the fact that all
 paths on the 2-tuple graph of {\bf A} are strongly simple, simple and
$SCC$-restricted.
  \medskip

  The determination of the order of local testability is in the general case
 NP-hard  $\cite {K94}$. But sometimes the situation is not so complicated.

\begin{thm} Let the state transition graph $\Gamma$ of a reduced deterministic
finite automaton be strongly connected.

Then the order of local testability of the automaton may be found in polynomial
time.
\end{thm}

  Proof. The verification of local testability is polynomial $\cite {K89}$.
 Finding the
graph $\Gamma\Gamma$ and its diameter is polynomial too. According to the
 preceding theorem it gives us the answer.
\medskip

\section{Necessary and sufficient conditions}

  In this section we assume that for every node ${\bf q} \in \Gamma$ and every
 element $\sigma \in \Sigma$ the node ${\bf q}\sigma$ exist (the transition
graph is complete). In general 
 it is not very strong
assumption because we can add to arbitrary graph $\Gamma$ a node ${\bf q}_0$
and suppose ${\bf q}\sigma={\bf q}_0$ in all undefined cases.

\begin{lem} $\label {4.1}$
Let $S$ be transition semigroup of a locally testable reduced
deterministic finite automaton. Let on the 2-tuple graph $\Gamma\Gamma$
of the automaton there are no $SCC$-restricted and $SCC$-semirestricted
paths of length $k-1$ and greater.

Let $x \in S^{k-1}, y,z \in S$ and $S$ satisfies identity
 $xyx=xyxyx$.

Then $S$ satisfies identity $xyxzx=xzxyx$ (Identity $\beta$ for
$k$-testability).
\end{lem}

Proof. Identity $xyx=xyxyx$ implies identities (4) and by
lemma $\ref {3.8}$ it implies the identity $xzxyx=xzxyxzx$.

  Let $\bf p$ be an arbitrary node of $\Gamma$.

  Consider the nodes ${\bf p}xzxz$ and ${\bf p}xyxzxz$. In case
${\bf p}xzxz={\bf p}xyxzxz$ we have
${\bf p}xzxyx={\bf p}xzxzxyx=={\bf p}xyxzxzxyx$ and from lemma
 $\ref {3.8}$ and identity $xyx=xyxyx$ it follows
that ${\bf p}xzxyx={\bf p}xyxzx$.  This implies that $xzxyx=xyxzx$.

So let us suppose that ${\bf p}xzxz \neq {\bf p}xyxzxz$.  Then the nodes ${\bf p}x$
and ${\bf p}xyxzx$ are distinct.

  Let us suppose that ${\bf p}xzxz \not\succ {\bf p}xyxzxz$. Consider
 the path $\phi$ from the node $({\bf p}, {\bf p}xyxzxz)$ to the node
  $({\bf p}x, {\bf p}xyxzxzx)=({\bf p}x, {\bf p}xyxzx)$ on $\Gamma\Gamma$.
 The length
of the path is not less than $|x| \geq k-1$. Note that the nodes ${\bf p}xzxz,
{\bf p}xyxzxz$ are reachable from the nodes ${\bf p}x$ and ${\bf p}xyxzx$
 by help
of the element $zxz$ and ${\bf p}xzxz \not\succ {\bf p}xyxzxz$.  Therefore
the path $\phi$ (or its part) is an $SCC$-semirestricted path of length $k-1$
or greater.  This contradicts the condition of lemma.

  So we may suppose that
${\bf p}xzxz \succ {\bf p}xyxzxz$
and ${\bf p}xyxy \succ {\bf p}xzxyxy$.
 Since the
nodes ${\bf p}xzxz$ and ${\bf p}xyxzxz$ have the common unit $xz$, from
  necessary
conditions of local testability (lemma $\ref {3.1a}$) it follows that the node
${\bf p}xyxzxzxyx={\bf p}xyxzxyx$ is reachable from the node
${\bf p}xzxzxyx={\bf p}xzxyx$.  In view of the lemma $\ref {3.8}$ we
  conclude that ${\bf p}xzxyx \succ {\bf p}xyxzx$.  From
  ${\bf p}xyxy \succ {\bf p}xzxyxy$
 it follows in analogous way that  ${\bf p}xyxzx \succ {\bf p}xzxyx$.

 So the nodes
${\bf p}xyxzx$ and ${\bf p}xzxyx$ belong to one $SCC$ of $\Gamma$. Then
  from (4) it
follows that the nodes ${\bf p}xyxzxz$ and ${\bf p}xzxyxy$ belong to the same
$SCC$.  The length of $x$ is not less then $k-1$ and is greater then the
length of every $SCC$-restricted path on $\Gamma\Gamma$.  So
${\bf p}xyxzxzx={\bf p}xzxyxyx$ and in view of $xyx=xyxyx$ we have
${\bf p}xyxzx={\bf p}xzxyx$ in this case too.

  Thus $xyxzx=xzxyx$.

 The lemma is proved.

\begin{lem} $\label {4.2}$
Let $k$ be a maximal number such that on the 2-tuple graph
$\Gamma\Gamma$ of deterministic finite locally testable reduced automaton
 {\bf A} there
exist $r$-periodic path of length $k+r$. Let $l$ be the maximum length
of $SCC$-restricted paths on $\Gamma\Gamma$. Let $m$ be the maximum
length of $SCC$-semirestricted paths on $\Gamma\Gamma$.
Let $n>max(k,l,m)+1$.

 Then {\bf A} is $n$-testable.
\end{lem}

  Proof. First consider the identities $\alpha_r$ of $n$-testability.
Let us suppose that for some elements $a_1,...,a_r$ from transition semigroup
$S$ of the automaton
\begin{equation}
(a_1...a_r)^{m+1}a_1...a_p \neq (a_1...a_r)^{m+2}a_1...a_p
\end{equation}
where $mr+p=n-1$, $ p<r$. Then for some node ${\bf q} \in \Gamma$ we have
${\bf q}(a_1...a_r)^{m+1}a_1...a_p \neq {\bf q}(a_1...a_r)^{m+2}a_1...a_p $.
Hence, on the graph $\Gamma\Gamma$ there exist $r$-periodic path from the node
(${\bf q}, {\bf q}a_1...a_r$) of the length $(m+1)r+p=(mr+p)+r$. In view of
equality $mr+p=n-1$ the length of the path is $n-1+r$. For $k=n-1$ we have
 $r$-periodical path of the length $k+r$. But it contradicts to our assumption
that $n>max(k,l,m) +1$ for all such $k$.
  So the identities $\alpha_r$ for $n$-testability hold in $S$.

  The validity of identity $\beta$ follows from the preceding lemma.

  The lemma is proved.
\medskip

  From the last lemma and lemmas $\ref {3.3}$, $\ref {3.4}$, $\ref {3.6}$
 follow now the necessary and sufficient conditions for the order of local
testability of deterministic finite reduced locally testable automaton.

 \begin{thm} $\label {4.3}$
 Let $k$ be the maximal natural number such that on the 2-tuple
 graph $\Gamma\Gamma$ of deterministic finite reduced locally testable
 automaton
 {\bf A} there exist $r$-periodic path of length $k+r$. Let $l$ be the maximum
 length of all $SCC$-restricted paths on $\Gamma\Gamma$. Let
$m$ be the maximum length of all $SCC$-semirestricted paths on
$\Gamma\Gamma$. Let $n=max(k,l,m)+2$.

  Then {\bf A} is precisely $n$-testable.
\end{thm}

\section{The upper bound}

\begin{lem}  $\label {5.1}$
Let $\Gamma\Gamma$ be the 2-tuple graph of locally testable
deterministic finite automaton with $n$ states.

 Then the length of any simple path
on the graph $\Gamma\Gamma$ is at most ${n^2-n \over 2}-1$.
\end{lem}.

  Proof. Any path on the graph $\Gamma\Gamma$could not contain both
 pairs $({\bf p},\bf q)$ and
 $(\bf q,\bf p)$ because it implies for some element $s$ of the transition
 semigroup that
$\bf q$$s={\bf p} $ and ${\bf p}s=\bf q $, whence some power of $s$ belongs to
 non-trivial group.
But locally testable semigroup do not contain non-trivial subgroups
  $\cite {BS}$.

 The number of non-ordered pairs with distinct components on an $n$-element
 set is equal to $n(n-1)/2$. Thus, the length of considered path is at most
$n(n-1)/2-1$.

 The lemma is proved.
 \medskip

\begin{thm}  $\label {5.2}$
Let $S$ be the transition semigroup of a locally testable
reduced deterministic
finite automaton with $n$ states.
 Then both $S$ and the automaton are $({n^2-n \over 2}+1)$-testable.
\end{thm}
  Proof immediately follows from theorem $\ref {3.9}$ and the preceding lemma.

\medskip

\section{Example for the upper bound}

   Let us consider the following example. Suppose the state transition graph
$\Gamma$ of the finite automaton {\bf M} contains $n$ nodes
 ${\bf q}_1,...,{\bf q}_n$, for $n>2$.
 Let
$\Sigma=\{ a, b_{i,j} \}$, where $i=1,...,n-2$, $n \geq j>i$.
 Suppose
${\bf q}_3a={\bf q}_1$. For $k \neq 3$ ${\bf q}_ka$
is undefined. Suppose
${\bf q}_ib_{i,j}={\bf q}_i$, ${\bf q}_jb_{i,j}={\bf q}_{j+1}$ for all $i,j$
such that $i<j<n $ and for
 $i<n-1$ ${\bf q}_ib_{i,n}={\bf q}_{i+1}$, ${\bf q}_nb_{i,n}={\bf q}_{i+2}$.
For other cases ${\bf q}_kb_{i,j}$ is undefined.

 It will be proved that the automaton {\bf M} is precisely
 $((n^2-n)/2+1)$-testable and so the upper bound of the order of testability
from theorem $\ref {5.2}$ is obtainable.

\begin{lem}  $\label {6.1}$
The state transition graph $\Gamma$ of
 the finite automaton {\bf M} is strongly connected. {\bf M} is locally
testable.
 \end{lem}

  Proof. In view of ${\bf q}_1={\bf q}_3a$, ${\bf q}_ib_{j,i}={\bf q}_{i+1}$
 and ${\bf q}_{n}b_{1,n}={\bf q}_3$
 the graph $\Gamma$
is strongly connected and all nodes of $\Gamma$ belong to one $SCC$.

  In  $\cite {K91}$ are given two conditions of local testability.  First is
the validity of the lemma 3.1 on $\Gamma$. Second must be verified only in
case $\Gamma$ is not an $SCC$. Thus according to lemma 3.1 we must prove
only that the distinct nodes of $\Gamma$ have no common unit in the
 transition semigroup $S$ of {\bf M}.

  Suppose ${\bf p}x=\bf p $, ${\bf q}x={\bf q} $, $\bf p \neq \bf q $ for
 $\bf p,{\bf q} \in \Gamma$, $x \in S$.
 Since there exists only one element of the kind ${\bf q}_ia$ the element
$x$ is not divided by $a$. So $x$ is a product of the $b_{i,j}$.

  From ${\bf p}x=\bf p \neq {\bf q}x=\bf q $ it follows that there is a
 cycle on
the 2-tuple graph $\Gamma\Gamma$ and all edges of the cycle are denoted by
$b_{i,j}$. Consider some node (${\bf q}_i,{\bf q}_j$) on the cycle. Suppose
 first $i>j$. Consider any existing node
(${\bf q}_i,{\bf q}_j$)$b_{l,r}=({\bf q}_{ii},{\bf q}_{jj})$. So $r=i$, $l=j$.
  We have either $({\bf q}_i,{\bf q}_j)b_{r,l}=({\bf q}_{i+1},{\bf q}_j)$ or
 in the case $i=n$
we have $({\bf q}_i,{\bf q}_j)b_{l,r}=({\bf q}_{j+2},{\bf q}_{j+1})$.
 Thus from $i>j$ it follows that
 $ii>jj$, $jj \geq j$ and in the case $jj=j$ we have $ii>i$ . So
$jj*n+ii>j*n+i$.

  Multiplication on $b_{l,r}$
induces a lexicographical order on the pairs ($\bf p,\bf q$) and all nodes
 on the path
with edges $b_{l,r}$ are distinct. So our assumption in the case $i>j$ is not
true.

  In the case $i<j$ we obtain contradiction too.

  Thus ${\bf p}x=\bf p $, ${\bf q}x=\bf q $ implies $\bf p=\bf q $. Therefore
 {\bf M} is locally testable.

\begin{lem} $\label {6.2}$
On the 2-tuple graph $\Gamma\Gamma$ of the automaton {\bf M}
there exists an $SCC$-restricted path of length ${n^2-n \over 2}-1$.
\end{lem}

 Proof. Consider the path:  (${\bf q}_1,{\bf q}_2$,), (${\bf q}_1,{\bf q}_3$),
...,(${\bf q}_1,{\bf q}_n$),(${\bf q}_2,{\bf q}_3$),...(${\bf q}_2,{\bf q}_n$),
...(${\bf q}_{n-2},\bf q_n$), ($\bf q_{n-1},{\bf q}_n$).
The nodes of the path are connected with edges noted by $b_{i,j}$. All nodes
of the kind (${\bf q}_i,{\bf q}_j$) such that $i<j$ belong to the path one
time.
The number of such nodes is ${n^2-n \over 2}$, so the length of the path is
${n^2-n \over 2}-1$. In view of the preceding lemma it is $SCC$-restricted
 path.

  \begin{thm} $\label {6.3}$
Deterministic finite automaton {\bf M} with $n$ states ($n>2$) is
 precisely (${n^2-n \over 2}+1$)-testable and its order of local testability
is equal to the upper bound on the order of local testability of a
deterministic finite reduced
automaton with  $n$ states.
  \end{thm}

 Proof. Lemma $\ref {6.1}$ gives us the local testability of {\bf M}.
 From theorem $\ref {5.2}$ follows
that for {\bf M} the upper bound of order of local testability is equal to
 $(n^2-n)/2+1$.
Lemma $\ref {3.4}$ in view of lemma $\ref {6.2}$ implies that the upper bound
 is reached on {\bf M}.

 The theorem is proved.
\medskip

 This implies the validity of the following statement
\begin{thm} The precise upper bound on the order of local testability of
deterministic finite locally testable reduced automata with $n$ states is
 equal to
${n^2-n \over 2}+1$
\end{thm}

  Proof. For $n>2$ it follows from the preceding theorem. For $n=2$ the
 semigroup of left zeroes gives us the needed example.
\medskip

\section{Example for two variables}

   Let us consider the following example of the  state transition graph
$\Gamma$ of the finite deterministic automaton ${\bf M}$:

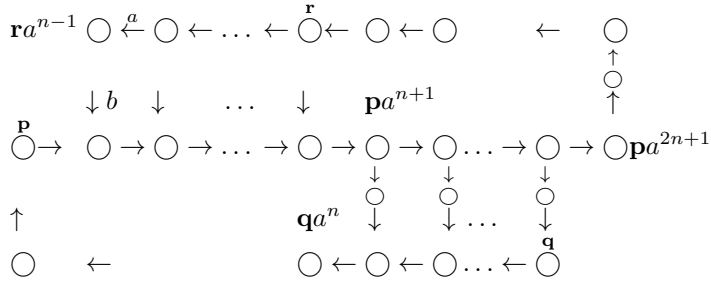
$$
\begin{array} {lllllll}
\hbox {${\bf r}a^{n-1}$}  &\bigcirc  \stackrel {a}{\leftarrow}
\bigcirc \leftarrow \ldots &\leftarrow &\stackrel
{{\bf r}} {\bigcirc} \leftarrow &\bigcirc \leftarrow \bigcirc & \leftarrow
& \bigcirc \\
 {} & \downarrow {b} \quad   \downarrow \qquad \ldots &&  \downarrow
 & \hbox{${\bf p}a^{n+1}$} && \stackrel
{\uparrow} {\stackrel {\bigcirc} {\uparrow}} \\

 \stackrel {{\bf p}} {\bigcirc}  \rightarrow &\bigcirc \rightarrow
 \bigcirc \to \ldots &
\rightarrow &\bigcirc \to &
\bigcirc  \to \bigcirc \ldots
 \to& \bigcirc  \to & \bigcirc \hbox {${\bf p}a^{2n+1}$} \\

  \uparrow &&& \hbox{${\bf q}a^n$}&
\stackrel {\downarrow}{\stackrel {\bigcirc} {\downarrow}}
    \qquad
 \stackrel {\downarrow}{\stackrel {\bigcirc} {\downarrow}}
 \ldots
&\stackrel {\downarrow} {\stackrel {\bigcirc} {\downarrow}} \\
  \bigcirc & \leftarrow   &
&\bigcirc  \leftarrow &\bigcirc \leftarrow \bigcirc
 \dots \leftarrow &\stackrel {{\bf q}} {\bigcirc}

\end{array}
$$
\medskip
 The vertical edges are noted by $b$, the horizontal edges are noted by $a$.

We have for $i<n$
 \[
 {\bf p}a^{2n}b^2={\bf q}, {\bf p}a^{n+1}b^2={\bf q}a^{n-1}, {\bf q}a^{n+1}b={\bf p}, {\bf p}a^{n+i}b^2={\bf q}a^{n-i}
\]
\[
{\bf r}b={\bf p}a^n, {\bf r}a^{n-i}b={\bf p}a^i,  {\bf r}a^{n-1}b={\bf p}a, {\bf p}a^{2n+1}b^2a^3={\bf r}
\]

So
\[
{\bf p}a={\bf p}a^{2n+1}b^2a^{n+2}b, {\bf p}={\bf p}a^{2n}b^2a^{n+1}b,
 {\bf p}={\bf p}a^{2n-i}b^2a^{n+1-i}b, {\bf p}a^{i}={\bf p}a^{2n+1}b^2a^{n+3-i}b.
\]

On the middle line there are $2n+2$ nodes, on the top line there are $n+3$
nodes, on the bottom line there are $n+2$ nodes.

 It will be proved that the order of local testability of the
automaton {\bf M} is $\Omega(n^2)$.

 Obvious is the following
\begin{lem} $\label {7.1}$
 The state transition graph $\Gamma$ of
 the finite deterministic automaton {\bf M} is strongly connected
 ({\bf M} is an $SCC$).
\end{lem}

\begin{lem} $\label {7.2}$
The finite deterministic automaton {\bf M} is locally
testable.
 \end{lem}

  Proof.
  In  $\cite {K91}$ are given two conditions for the automaton to be
locally testable.
First is the validity of lemma 3.1 on $\Gamma$. Second must be verified only
in the case that $\Gamma$ is not an $SCC$. Thus, by the preceding lemma, we
 must prove only
 that distinct nodes of $\Gamma$ have no common unit in transition semigroup
$S$ of {\bf M}.

  Suppose that there are two cycles on $\Gamma$ with edges corresponding
the element $x=a^{k_1}b^{l_1}a^{k_2}b^{l_2}...a^{k_s}b^{l_s}$ from transition
semigroup $S$ of {\bf M}. Our aim is to prove that both the cycles coincide.

Let us assume that for nodes ${\bf f}, {\bf g} \in \Gamma$ we have
 ${\bf f}x={\bf f}, {\bf g}x={\bf g}, {\bf f} \neq {\bf g}$.

 Let $x_i$ be the left subword of the word $x$ of length $i$.

Then ${\bf f}x_i \neq {\bf g}x_i$ for any $i$ and there exist a cycle in
the 2-tuple graph
$\Gamma\Gamma$ with the nodes $({\bf f}x_i, {\bf q}x_i)$, $0<i \leq |x|$.

 It is not difficult to see that $l_j=1$ or $l_j=2$ and $l_j+l_{j+1}=3$.
Without loss of generality we can assume that $l_1=2$.
The nodes ${\bf f}a^{k_1}b^2={\bf f}x_{k_1+2}$ and ${\bf g}a^{k_1}b^2={\bf g}x_{k_1+2}$ exist only if
${\bf f}x_{k_1}={\bf p}a^m$ and ${\bf g}x_{k_1}={\bf p}a^l$, for some $m,l>n$.
Both the nodes ${\bf f}a^{k_1}b^2a^{k_2}b$ and ${\bf g}a^{k_1}b^2a^{k_2}b$
exist and
are distinct. They belong to the middle line and are presented in the form
${\bf p}a^i$ ($i \geq 0$).
Since not more than one of them may be ${\bf p}$, another is equal to
 ${\bf p}a^{i}$ where $i \geq 1$.
 Let us suppose that
${\bf f}a^{k_1}b^2a^{k_2}b={\bf p}a^i$ for $i>0$. Then
  ${\bf f}a^{k_1}b^2={\bf p}a^{2n+1}$
 and ${\bf g}a^{k_1}b^2={\bf p}a^{n+t}$ where $0<t<n+1$.
So to the cycle of $\Gamma\Gamma$ belongs the node
$({\bf p}a^{2n+1}, {\bf p}a^{n+t})$.
Then the node $({\bf p}a^{2n+1}b^2, {\bf p}a^{n+t}b^2)$ belongs to the
 same cycle.
It implies that the node
 $({\bf p}a^{2n+1}b^2a^3, {\bf p}a^{n+t}b^2a^3)=({\bf r},{\bf q}a^{n-t+3})$ is
 on the same cycle too.
The second component of one of the nodes on considered cycle of
$\Gamma\Gamma$ must to be ${\bf p}$.
From ${\bf q}a^{n-t+3}a^jb={\bf p}$ follows that
$j+n-t+3=n+1$ and $j=t-2$. Then the first component of the same node
is ${\bf r}a^{j}b={\bf p}a^{n-j}$.
From the node $({\bf p}a^{2n+1}, {\bf p}a^{n+t})$ we reach the node
 $({\bf p}a^{n-t+2}, {\bf p})$
and therefore the node
$({\bf p}a^{2n+1}, {\bf p}a^{n+t-1})$.

Distance between components of the nodes is growing from $a^{n-t+1}$
to $a^{n-t+2}$. So for subword of $x$ containing two distinct inclusions of $b$
 ($b^2$ and then $b$)
distance between components is growing. Obvious that $s$
 is even number. So the distance between two components of the node is growing
on the path corresponding $x$.
This contradicts to the fact that $x$ defines
 the cycle
on $\Gamma\Gamma$ (Or two distinct corresponding cycles on $\Gamma$).

 So {\bf M} is locally testable.

\begin{lem} $\label {7.3}$
 On the 2-tuple graph $\Gamma\Gamma$ of the automaton {\bf M}
there exist a $SCC$-restricted path of length $2n^2+4n-6$.
\end{lem}

 Proof. Consider the path defined by the word
$a^{2n+1-i}b^2a^{n+2-i}b$ from the node
 $({\bf p},{\bf p}a^i)$ for $0<i<n$. We have
\[
({\bf p},{\bf p}a^i)a^{2n+1-i}b^2a^{n+2-i}b=({\bf p},{\bf p}a^{i+1}).
\]

 The length of the path is equal to $3n+6-2i$,
the final node is $({\bf p},{\bf p}a^{i+1})$.

  Now consider the sequence of such paths for $i=1,2,...,n-1$.
We get a path from the node $({\bf p},{\bf p}a)$ to the node
$({\bf p},{\bf p}a^{n})$.
 The length of the path is $2n^2+4n-6$.

 The lemma is proved.

  \begin{thm} Deterministic finite automaton {\bf M} whose alphabet
size is two is locally testable and its order of local testability
is $\Omega(n^2)$.
  \end{thm}

 Proof. Lemma $\ref {7.2}$ gives us the local testability of {\bf M}.
Number of nodes in {\bf M} is equal to $5n+8$ and is linear in $n$.
According the preceding lemma there exist a path of length $2n^2+4n-6$ on the
 2-tuple graph of the
automaton. In view of lemma $\ref {3.4}$, this number
gives us a lower bound for the order of local testability.

So the lower bound for the order of local testability is $\Omega(n^2)$.
According theorem $\ref {5.2}$ (see $\cite {K91}$ too) it is an upper bound
as well.
 \medskip

\section{Acknowledgments}

  The author thanks professor Margolis for helpful discussion and for many
useful suggestions.


\begin{thebibliography}{99}
  \bibitem{BS} J.A. Brzozowsky, I. Simon, Characterization of locally testable
     events, Discrete Math., 4, (1973) 243-271.
\bibitem{G} A. Ginzburg, About some properties of definite, reverse-definite
and related automata, IEEE Trans. Electron. Comput. ES15(1966) 806-810.
\bibitem{C} P. Caron, AG: Families of locally testable languages.
Rapport LIR 97.03.Univ. de Rouen,Fr. 1997, 206-210.
\bibitem{K89} S. Kim, R. McNaughton, R. McCloskey, An upper bound on the
order of locally testable deterministic finite automaton, Lect.  Notes in
  Comp., 401, (1989) 48-65.
\bibitem{K91} S. Kim, R. McNaughton, R.
 McCloskey, A polynomial time algorithm for the local testability problem of
deterministic finite automata, IEEE Trans. Comput. 40(1991) N10, 1087-1093.
\bibitem{K94} S. Kim, R. McNaughton, Computing the order of locally
   testable automaton, SIAM J. Comput., 23(1994), 1193-1215.
\bibitem{L} G. Lallement, Semigroups and combinatorial applications, Wiley,
N.Y., 1979
\bibitem{Ma} S. W. Margolis, J.E.Pin. Languages and inverse semigroups,
 11 ICALP, Lect. Notes in
  Comp. Sc, 199, Springer, Berlin(1985) 285-299.
 \bibitem{MP} R. McNaughton, S. Papert, Counter-free automata,
  M.I.T.  Press Mass., 1971.
\bibitem{Mi} M. Minsky, S. Papert, Perceptrons,
M.I.T. Press Mass., 1971, Cambridge, MA, 1969.
\bibitem{Pa} Pastijn, F., Regular locally testable semigroups as semigroups of
quasi-ideals, Acta Math. Hung., 36, 1-2(1980), pp. 161-166.
\bibitem{Pi} Pin J., Finite semigroups and recognizable languages. An
introduction. Semigroups and formal languages, Math. and Ph. Sc., v.466,
(1995), pp. 1-32.
 \bibitem{Sh} M. Perles, M. O. Rabin, E. Shamir, The theory
 of definite automata, IEEE Trans. Electron. Comput. ES-12(1963) 233-243.
\bibitem{Tr} A.N. Trahtman,  The varieties of testable semigroups.
   Semigroup  Forum, 27, (1983), 309-318.
\bibitem{TW} A.N. Trahtman, Precise upper bound on the order of local
 testability of finite automaton. 2 Int. Workshop on Impl.
Automata, Univ. of Western Ontario, Canada, 1997, 113-121.
 \bibitem{Z} Y. Zalcstein,
Locally testable semigroups, Semigroup Forum, 5, (1973), 216-227.
\bibitem{ZC} Zalcstein, Y., Syntactic semigroups of some classes
 of star-free languages, J. Comp. System Sci., 6, (1972), pp. 151-167.

  \end{thebibliography}
 \end{document}